\input harvmac
\noblackbox
\def\ias{\vbox{\sl\centerline{School of Natural Sciences, 
Institute for Advanced Study}%
\centerline{Olden Lane, Princeton, N.J. 08540 USA}}}

\lref\one{C. Hull and P. Townsend, Nucl. Phys. {\bf{B438}} (1995) 109,
hep-th/9505073.}
\lref\two{E. Witten, Nucl. Phys. {\bf{B443}} (1995) 85, hep-th/9503124.}
\lref\three{A. Sen, Int. J. Mod. Phys. {\bf A9} (1994) 3707, hep-th/9402002.}
\lref\four{J. Blum, Phys. Lett.  {\bf{B333}} (1994) 92, hep-th/9401133.}
\lref\five{A. Sen, Phys. Lett. {\bf{B329}} (1994) 217, hep-th/9402032.}
\lref\mir{M. Cveti\v{c} and D. Youm,  Phys. Lett. {\bf{B359}} (1995) 87,
hep-th/9507160.}
\lref\six{E. Witten, Nucl. Phys. {\bf{B460}} (1996) 541, hep-th/9511030.}
\lref\seven{S. Sethi and M. Stern, hep-th/9607145.}
\lref\eight{E. Witten, Nucl. Phys. {\bf{B202}} (1982) 253.}
\lref\nine{E. Witten, Nucl. Phys. {\bf{B460}} (1996) 335, hep-th/9510135.}
\lref\ten{A. Sen, Phys. Rev. D54 (1996) 2964, hep-th/9510229.}
\lref\eleven{A. Sen, Phys. Rev. D53 (1996) 2874, hep-th/9511026.}
\lref\twelve{M. Atiyah and N. Hitchin, {\it The Geometry and Dynamics of
Magnetic Monopoles}, Princeton University Press (1988).}
\lref\por{M. Porrati, Phys. Lett. {\bf{B377}} (1996) 67, hep-th/9505187.}
\lref\porh{M. Porrati, Phys. Lett. {\bf{B387}} (1996) 492, hep-th/9607082.}
\lref\seg{G. Segal and A. Selby, Comm. Math. Phys. 177 (1996) 775.}
\lref\dh{A. Dabholkar and J. Harvey, Phys. Rev. Lett. 63 (1989) 498\semi
A. Dabholkar, G. Gibbons, J. Harvey, and F. Ruiz-Ruiz, Nucl. Phys. 
{\bf{B340}} (1990) 33.}
\lref\pdl{J. Dai, R. Leigh, and J. Polchinski; Mod. Phys. Lett. A4 (1989) 273.}
\lref\ber{M. Berkooz {\it et al.}, Nucl. Phys. {\bf{B475}} (1996) 115, 
hep-th/9605184.}

\Title{\vbox{\baselineskip12pt
\hbox{IASSNS-HEP-97/8}\hbox{hep-th/9702084}}}
{\vbox{\centerline{H-Dyons and S-Duality}}}

{\bigskip
\centerline{Julie D. Blum}
\bigskip
\ias

\bigskip
\medskip
\centerline{\bf Abstract}

We present a relatively simple argument showing that the 
H-dyon states required by
S-duality of the heterotic string on $T^6$ are present provided that
the BPS dyons required by S-duality of $N=4$ supersymmetric
Yang-Mills theory are present.  We also conjecture and provide evidence that
H-dyons at singularities where the nonperturbative gauge symmetry is
completely broken are actually BPS dyons.} 

\Date{2/97}

\newsec{Introduction}

The known superstring theories have all been conjectured to be related 
to each other through dualities acting on the coupling constant and the 
target space \refs{\one , \two}.  In other words, spacetime dualities are
in some sense equivalent to worldsheet dualities.  Evidence has
mounted during the last several years that these conjectures are correct.  The
first conjectured weak-strong coupling duality of superstrings
involved the heterotic string compactified on a six-torus.  This
theory is believed to have an $SL(2,{\bf{Z}})$ symmetry (S-duality)
acting on the complex coupling as well as the electric and magnetic
excitations.  Supporting evidence and predictions of heterotic string 
S-duality were presented in Ref.~\three .  One of these predictions
concerns the existence of the so called ``H-dyon'' states.  The purpose
of this note is to argue that these states exist and to reveal their
whereabouts.

If S-duality is truly a symmetry of the heterotic string on a six-torus,
all measurable quantities must be invariant under $SL(2,{\bf{Z}})$ 
transformations.  These transformations convert elementary electrically
charged states into states with both electric and magnetic charges
(dyons) at a new value of the coupling.  In some cases these dyon
states cannot decay into any lower energy states that conserve
their charges and are expected to be stable states.  S-duality
predicts that their degeneracies, in these cases, should be equal to
those of the corresponding elementary states.  The H-dyon states to
be discussed here are the S-duality transforms of a certain class
of elementary string states to be specified in section two.  In that
section we will refresh the reader's memory about some facts related 
to S-duality.  Under the special weak-strong coupling transformation
of $SL(2,{\bf{Z}})$, the above class of elementary string states
become the H-monopoles.  The evidence that H-monopole degeneracies
agree with the predictions of S-duality is discussed in section three.
In section four we discover the H-dyons according to the prediction
so long as another class of dyons, the BPS dyons, is also detected.
We summarize the results in section five.  Along the way we conjecture that 
H-monopoles or H-dyons at singularities where the nonperturbative
gauge symmetry is broken completely but the gauge group is abelian
are actually BPS monopoles or BPS dyons.

\newsec{Review of S-Duality}

To set the stage for our argument let us review the background details
about S-duality of the heterotic string.  The discussion will follow 
Ref.~\three.  Following the convention that the right moving sector is 
supersymmetric whereas the gauge symmetry resides in the left moving
sector, the masses of perturbative string states can be written in
the Neveu-Schwarz sector by the following relation for a particular
choice of background values on the six-torus of the metric, 
antisymmetric tensor,
and sixteen $U(1)$ gauge fields from the ten dimensional gauge group
$SO(32)$ or $E_8\times E_8$:  
\eqn\mass{M^2\propto{1\over {(Im\tau)\alpha '}}({\vec{p}_R}^{\,2}+2N_R-1)=
  {1\over {(Im\tau)\alpha '}}({\vec{p}_L}^{\,2}+2N_L-2).}
The Ramond sector masses are degenerate by supersymmetry.
The left and right internal momenta and winding vectors 
$(\vec{p}_L,\vec{p}_R)\in\Lambda_{22,6}$
where $\Lambda_{22,6}$ is an even, self-dual Lorentzian lattice
also determined by the position in the moduli space.
The $N_L,\, N_R$ are left and right moving oscillator numbers, 
$-1/2$ and $-1$ are the right and left moving vacuum energies, and
$\tau$ is the asymptotic value of the complex coupling  
($\tau={\theta\over{2\pi}}+{i\over{g^2}}$; $\theta$ is the axion,
$g^2$ is the string loop expansion parameter).  The string tension is
$T=1/2\pi\alpha '$.

It is known that the $N=4$ supersymmetry of the heterotic
string on a six-torus protects the masses of states that satisfy a
Bogomol'nyi bound from receiving quantum corrections.  The
elementary electrically charged string states breaking half of the
supersymmetry and saturating this bound have $N_R=1/2$
but are otherwise arbitrary.  These states
satisfy the 
relation: $N_L -1={1\over 2}({\vec{p}_R}^{\,2} -{\vec{p}_L}^{\,2})$.
There are also states not seen in the perturbative string theory
that contain both electric and magnetic charges.
One can write these charges in the form
\eqn\charge{(\vec{Q}_{el},\vec{Q}_{mag})=({1\over{Im\tau}}(\vec{p}+Re\tau
\vec{q}\,),\vec{Lq}\,)}
where $(\vec{p},\vec{q}\,)\in\Lambda_{22,6}$, $L$ is the $28\times 28$
matrix with 
\eqn\L{L=\left(\matrix{0&I_6&0\cr I_6&0&0\cr0&0&-I_{16}\cr}\right),}
 and $I_n$ is the $n\times n$ identity matrix.  One can also write
the left and right moving vectors as projections onto the subspaces
$L=\mp 1$:
\eqn\project{\eqalign{p_L^a&={1\over 2}(I_{28}-L)_{ab}p^b\cr
p_R^a&={1\over 2}(I_{28}+L)_{ab}p^b\cr}.}
The first six components of $Q_{el}$ are charges with respect to
the gauge fields from the dimensional reduction of the ten dimensional
metric on $T^6$, the next six are charges with respect to the gauge
fields from the antisymmetric tensor, and the last sixteen are charges
of the sixteen $U(1)$ fields.  Under the $SL(2,{\bf{Z}})$ transformation
$$\left(\matrix{a&b\cr c&d\cr}\right),$$
the coupling and lattice vectors transform as follows:
\eqn\trf{\eqalign{\tau&\rightarrow{a\tau+b\over{c\tau+d}}\cr
\noalign{\hbox{}}
 \left(\matrix{\vec{p}\cr\vec{q}\cr}\right)&\rightarrow
\left(\matrix{a&-b\cr -c&d\cr}\right)\left(\matrix{\vec{p}\cr\vec{q}\cr}
\right)\cr}.}
  The weak-strong coupling transformation corresponds to the
 $SL(2,{\bf{Z}})$ matrix 
$$\left(\matrix{0&1\cr -1&0\cr}\right).$$
The mass formula can be written in a manifestly $SL(2,{\bf{Z}})$ invariant
fashion.  One would like to show that the spectrum of charges is
also invariant under $SL(2,{\bf{Z}})$.  One must, therefore, determine
the degeneracies of various dyonic states.  In general, an elementary
string state
$\vec{p}=\vec{e}$ where $\vec{e}$ is an elementary lattice vector
transforms into a state $m\vec{e}+n\vec{\tilde{e}}$ where $\vec{\tilde{e}}$
is a dual lattice vector (also a lattice vector in this case).  In the
case that ($m,n$) are relatively prime, the mass formula implies
that the corresponding state should be stable, and $SL(2,{\bf{Z}})$
invariance implies that the degeneracy should equal the corresponding
degeneracy of elementary states.
 
The sixteen states (eight Neveu-Schwarz states and eight Ramond
states from the right moving sector)
with ${\vec{p}_L}^{\,2} -{\vec{p}_R}^{\,2}=2$ $(N_L=0)$ 
which are electrically charged
under one of the sixteen $U(1)$'s have an
interpretation as elementary charged particles arising from an
$N=4$ supersymmetric Yang-Mills theory in four dimensions in which the
gauge group is broken to
an abelian subgroup by expectation values of the scalar fields.
Since the masses of these particles can vanish in the field theory limit, 
gravity can be ignored.
S-duality changes these states into BPS dyons.  
It was shown in Ref.~\four\ that for the $SU(2)$ case, the theory
of the sixteen magnetically charged states resulting from a weak-strong
coupling $SL(2,{\bf{Z}})$ transformation at low energies and weak
coupling is an $N=4$ supersymmetric quantum mechanics on the moduli
space of classical monopole solutions as expected by S-duality.  It
was conjectured \five\ that there should be unique harmonic forms on this
moduli space for each $(n,m)$ with $n,m$ relatively prime integers 
and $n$ the number of magnetic charges and $m$ an integer such that
the electric charge is proportional to $m+{n\theta\over{2\pi}}$.

There are $24\times 16$ states with ${\vec{p}_L}^{\,2} -{\vec{p}_R}^{\,2}=0$
since there are $24$ left moving oscillators  with $N_L =1$ ($8$ space-time
oscillators and $16$ $U(1)$ oscillators). 
These states arise as
momentum or winding states on the six-torus.  They are electrically
charged with respect to a gauge field coming from the metric or
antisymmetric tensor.  It was conjectured \three\ that there should be
$16\times 24$ H-dyon bound states satisfying the Bogolmol'nyi bound
for each $(n,m)$ with $n,m$ relatively
prime integers with $n$ the instanton number and
$m$ the momentum on $S^1$.
Under the weak-strong coupling transformation
the momentum states on an $S^1$ turn into magnetically charged states
with respect to the gauge field from the antisymmetric tensor.  They
satisfy $dH=F\wedge F$ corresponding to one instanton lying 
in $R^3\times S^1$,
and they are known as H-monopoles (H is the antisymmetric tensor field
strength, F is the gauge field strength).  The solitonic 
states with $N_L>1$
are always heavy in the field theory limit and can never be constructed
from massless string fields as have been, for example, certain dyons with
$N_L=1$
at special values of the moduli \mir .  To reiterate, our goal is to
locate the $16\times 24$ $(n,m)$ H-dyon bound states when $n$ and $m$
are relatively prime integers.

\newsec{H-Monopoles Unmasked}

The expected H-monopoles corresponding to $n=1$ and $m=0$ 
were shown to exist in Refs.~\refs{\six ,\seven}.  (See also Ref.~\porh.)  
The $E_8\times E_8$
heterotic string on $T^6$ is indistinguishable from the $SO(32)$
one because of T-duality so we may as well consider the H-monopole
as an instanton of $SO(32)$.
Generically, the $SO(32)$ gauge symmetry is broken to $U(1)^{16}$ by 
Wilson lines on $T^6$, and the instanton shrinks to zero size \six .
Zero size instantons can occur even when the gauge symmetry is nonabelian.
The moduli space of the relevant small instantons was described 
\six\ using heterotic-type I duality with the small instanton
being dual to a Dirichlet fivebrane wrapped around the five-torus transverse
to $R^3\times S^1$.  The D5 brane of the type I theory carries an 
$Sp(1)$ vector, a neutral hypermultiplet, and a $(2,32)$ half hypermultiplet
of $Sp(1)\times SO(32)$ at maximal symmetry points in the moduli space.
Symmetry breaking patterns can be
determined by adding Wilson lines of $Sp(1)$ to $T^5$ and $SO(32)$ to
$T^6$, alternatively by giving expectation values to the charged
matter, or by some combination of the two.  From the point of view
of the heterotic string, the $Sp(1)$ symmetry is nonperturbative, and
the instanton classically remains small so long as this 
symmetry is not completely
broken.  Finite size instantons have no enhanced gauge symmetry.
The symmetry can be completely broken at singularities in the
moduli space of Wilson lines where the interaction of $SO(32)$ and
$Sp(1)$ Wilson lines cause a charged hypermultiplet to become
massless.  Yet in this case, assuming the perturbative gauge
symmetry is $U(1)^{16}$, the instanton must classically remain small.  I
would like to conjecture that in the quantum theory 
a small instanton at such a singularity
is equivalent to a BPS monopole (certainly the two moduli spaces
are in agreement and the nonperturbative gauge symmetry is broken).

Experience tells us that if the singularity at the location of the
small instanton were not smoothed, we should expect extra degrees of
freedom to enter there.  Heuristically, one can understand that at
the singularity a $U(1)$ of the $SO(32)$ becomes identified with the
$U(1)$ of the broken $Sp(1)$, and one can identify the six Wilson lines in 
this $U(1)$ with the six neutral scalars of the $N=4$ Yang-Mills
theory in four dimensions with gauge group $SU(2)$.  The monopole
is an $Sp(1)$ monopole that breaks half the supersymmetry on the brane
($N=4$ in four dimensions is broken to $N=2$ at the singularity).
If we T-dualize on the $p$ circles with $Sp(1)$ Wilson lines, we
obtain a $5-p$ brane that touches a $9-p$ brane at the singularity
so the dependence on separations in the $p$ dimensions has 
disappeared at this point as expected for a monopole solution.

In Ref.~\six\ the moduli space of H-monopoles was considered at points
where the $SO(32)$ was broken to $U(1)^{16}$ by Wilson lines and
the $Sp(1)$ was broken to $U(1)$ in the same way.  At generic points
the theory was noninteracting modulo the Weyl group, 
and sixteen states came from quantizing the 
zero modes of the hypermultiplet.  The vector only gave eight states
because it was necessary to divide by the Weyl symmetry of 
$Sp(1)\,({\bf{Z_2}})$.  The total degeneracy was, therefore, $16\times 8$
states.

The remaining $16\times 16$ states were conjectured \six\ to come
from the sixteen singularities where the two types of Wilson lines
interacted such that a charged hypermultiplet became massless.
That a normalizable supersymmetric state comes from each singularity
was proved by Ref.~\seven\ (see also Ref.~\porh) by reducing the gauge 
theory near the 
singularity to a quantum mechanics and computing the index \eight\ of
ground states weighted by $(-1)^{F}$ where $F$ is the fermion
number.  If our conjecture is correct, this result
follows from the existence of the BPS monopole.

\newsec{H-Dyons Revealed}

We would like to prove that the expected H-dyon states are present for
$n,m$ relatively prime and $m\not= 0$.  Part of
our argument will rely on the existence of the corresponding BPS dyon
bound state.  Again we will have two contributions to the H-dyon states
from generic points in the moduli space and from singularities.
Let us summarize our strategy.  We will first discuss the type I
formulation of the generic H-dyons.  There we will see that the 
number of center of mass states is $16\times 8$.  In order to
determine whether a bound state exists, we will find it easiest to 
T-dualize to a type I' system and then dimensionally reduce the 
system to two dimensions where the supersymmetry will be $N=8$.
At this point we will imitate the arguments of Ref.~\nine\ to show
that there are $16\times 8$ generic bound states.  We will then
argue as in Ref.~\six\ that the remaining H-dyon states come from
sixteen singularities where the supersymmetry is halved.  Using
arguments from Refs.~\refs{\nine,\eleven,\twelve}, we will show that at these
singularities the system of H-dyons can be considered, for the purposes
of determining the degeneracy of supersymmetric bound states, as an
$N=4$ supersymmetric quantum mechanics on the moduli space of BPS dyons.

\subsec{Generic H-Dyons}

We begin with the type I formulation of the H-dyons.
Let us assume again that $SO(32)$ is broken to $U(1)^{16}$.
The maximal gauge symmetry on $n$ coincident fivebranes 
is $Sp(n)$ with matter in a reducible antisymmetric tensor and $16$
fundamentals.  In addition to $SO(32)$ Wilson lines, we want to include
a special $Sp(n)$ Wilson line that takes values in one $U(1)$ breaking
the gauge symmetry to $SU(n)\times U(1)$ with an adjoint hypermultiplet.  The
fundamentals and antisymmetric tensors of $SU(n)$ are charged under the
Wilson line $U(1)$ so only the adjoint remains massless.  Such a
Wilson line $W^{\rho}$ in the fundamental $2n$ representation can take the form
\eqn\Wil{W^{\rho}=A^{\rho}\otimes I_n}
with
\eqn\A{A^{\rho} (\beta^{\rho})=\left(\matrix{\cos\beta^{\rho} &
-\sin\beta^{\rho}\cr 
\sin\beta^{\rho} &\cos\beta^{\rho}\cr}\right)}
where $A^{\rho}({\pi\over 2})\otimes I_n$ is preserved by $Sp(n)$ and 
${\rho}$ indexes
one of the five $S^1$'s.  Notice that the
symmetry is broken to $SU(n)\times U(1)$ for any $\beta^{\rho}\neq 0,\pi$ 
and that
$Sp(k)$ symmetry ($1\leq k\leq n$) is only achieved on the boundary of 
the moduli space.
There is a subtlety in this problem related to the fact that the Weyl
group of $Sp(n)$ is ${\bf{Z_2}}^n\times S_n$ rather than $S_n$ which is
the Weyl group of $SU(n)$ ($S_n$ is the symmetric group on $n$ elements).
In general, we can further break the gauge symmetry to $U(1)^n$ by a
Wilson line of the form $A^{\rho} (\beta_1^{\rho})\oplus A^{\rho} 
(\beta_2^{\rho})\oplus ... 
\oplus A^{\rho} (\beta_n^{\rho})$.  
The center of mass vector will be located at 
$\beta^{\rho}={1\over n}
(\beta_1^{\rho}+...+\beta_n^{\rho})$ which is invariant under $S_n$.  Only an
overall ${\bf Z_2}$ commutes with this $S_n$.  A general Weyl transformation
will take this center of mass vector multiplet into the center of mass
multiplet in another fundamental region of ${\bf{Z_2}}^{n-1}$.  Thus,
we can restrict ourselves to the fundamental region of ${\bf{Z_2}}^{n-1}$
with $0\leq\beta^{\rho}_j\leq\pi$ or $-\pi\leq\beta^{\rho}_j\leq 0$.  
In a fundamental region, we can 
consider the symmetry breaking as follows:
\eqn\sym{\eqalign{Sp(n)&\rightarrow SU(n)\times U(1)\cr
n(2n+1)&\rightarrow n(1)+{\bar n}(-1)+{n(n-1)\over 2}(1)+
{{\bar n}({\bar n}-1)\over 2}(-1)+(n{\bar n}-1)(0)+1(0)}}

This system
has $N=2$ supersymmetry in six dimensions, away from singularities
where charged hypermultiplets become massless or $Sp(k)$ symmetry 
($1\leq k\leq n$) is restored such that the supersymmetry is
reduced to $N=1$.
Next we T-dualize on the $S^1$ transverse
to the fivebrane and dimensionally reduce the resulting type I' theory
to the $S^1/{\bf Z_2}$ plus time.  We will show that the number of 
center of mass degrees of
freedom of the ground states has not changed by the T-duality or
dimensional reduction, and it is easiest to count in six dimensions
(on the fivebrane) before T-dualizing.
The center of mass hypermultiplet is invariant under the Weyl group, and the
quantization of fermion zero modes provides sixteen states.  The center
of mass vector multiplet provides eight states after dividing by ${\bf Z_2}$.
The total number of center of mass states is $16\times 8$.

There are two 
equivalent pictures of the H-dyons.  We can consider them as fivebranes
that carry momentum on $S^1$ or we can consider them as fivebranes 
interacting with elementary charged states of the string.
The T-duality will convert the type I theory on $T^5\times S^1$
into a type I' theory on $T^5\times S^1/{\bf Z_2}$.  In the second 
picture the $n$ fivebranes become $2n$ sixbranes wrapped on $S^1/{\bf Z_2}$,
and the elementary charged states of momentum $m$ become elementary winding
states of winding number $m$.  At the two orientifold planes the
gauge theory is the same $Sp(n)$ gauge theory that we have discussed
in the type I picture.  Away from these planes the theory is locally a
type IIA theory with $U(2n)$ gauge symmetry \pdl .  If we restrict 
ourselves to $0\leq\theta\leq\pi$ where $\theta$ is the angle on
$S^1$, the winding can be considered equivalent to putting the
tensor product of $m$ quarks in the fundamental $\bf{2n}$ representation
of $U(2n)$ at $\theta=0$ and the tensor product of $m$ quarks in
the fundamental $\bf{\overline{2n}}$ of $U(2n)$ at $\theta=\pi$ \nine .  
The ${\bf Z_2}$
action takes $\theta$ to $-\theta$ and the $\bf{2n}$ to the 
$\bf{{\overline{2n}}}$.
There is, thus, an electric flux between the two orientifold planes
which vanishes at the two planes.  Thus, we
can reduce the problem to counting the number of supersymmetric 
ground states of the gauge theory on the brane with an electric
flux around the $S^1/{\bf Z_2}$.  Since there are no momentum or winding states
on the five circles transverse to the $S^1/{\bf Z_2}$ 
and the problem here is to
count the ground states, we can dimensionally reduce the gauge theory
on the brane to the $S^1/{\bf Z_2}$ plus time.  
The $N=2$ supersymmetry in six dimensions reduces to
$N=8$ in two dimensions away from the above mentioned singularities.

Now let us try to understand in detail the gauge theory that we must 
analyze.  We will choose a representation of $U(2n)$ on the 
covering space of the orbifold with $4n$ sixbranes such that an
element of the Lie algebra of $U(2n)$ is written as follows:
\eqn\u{U=\left(\matrix{S_{2n}&A_{2n}\cr A_{2n}&S_{2n}\cr}\right).}
The $2n\times 2n$ dimensional matrices $S_{2n}$ and $A_{2n}$ can be
decomposed as follows:
\eqn\dec{\eqalign{S_{2n}&=\left(\matrix{A_n +iS_n&S_n^{'} +iS_n^{''}\cr
-S_n^{'} +iS_n^{''}&A_n -iS_n\cr}\right)\cr
A_{2n}&=\left(\matrix{A_n^{'} +iS_n^3&A_n^{''}+iA_n^3\cr
A_n^{''} -iA_n^3&-A_n^{'} +iS_n^3\cr}\right)\cr}}
where $A_n$, $A_n^{'}$, $A_n^{''}$, and $A_n^3$ are $n\times n$
dimensional antisymmetric matrices; and $S_n$, $S_n^{'}$, $S_n^{''}$, 
and $S_n^3$ are $n\times n$ dimensional symmetric matrices.  The
gauge group $Sp(n)$ is generated by $Re S_{2n}$ and $Im S_{2n}$.
This $Sp(n)$ preserves the metric 
\eqn\met{G_{2n}=\left(\matrix{0&I_n\cr -I_n&0\cr}\right).}
The ${\bf Z_2}$ monodromy is embedded in $U(2n)$ by the matrix
\eqn\m{M=\left(\matrix{I_{2n}&0\cr 0&-I_{2n}\cr}\right).}
This monodromy has the following action:
\eqn\act{\eqalign{S_{2n}&\rightarrow S_{2n}\cr A_{2n}&\rightarrow
-A_{2n}}.}
At the orientifold planes, $A_{2n}$ is projected out, and the remaining
gauge symmetry is $Sp(n)$.  The $2n$ $U(1)$'s of $U(2n)$ will be the
following $(1\leq i\leq n)$ :
\eqn\abe{\eqalign{G_{\beta_i}&=I_2\otimes g_{\beta_i}\cr
G_{\alpha_i}&=\left(\matrix{0&g_{\alpha_i}\cr g_{\alpha_i}&0\cr}\right)
\cr}}
where $g_{\beta_i}$ and $g_{\alpha_i}$ are $2n\times 2n$ dimensional
matrices such that
\eqn\g{\eqalign{(g_{\beta_i})_{jk}&=\delta_{j,i}\delta_{k,n+i}-
\delta_{j,n+i}\delta_{k,i}\cr
(g_{\alpha_i})_{jk}&=i\delta_{j,i}\delta_{k,i}+i\delta_{j,n+i}\delta_{k,n+i}
\cr}.}
The $G_{\beta_i}$'s
lie in an $Sp(n)$ subgroup.

If we denote the Wilson lines, parametrized by angles $\alpha_i$ and
$\beta_i$, corresponding to the above $U(1)$'s as $W_{\alpha_i}$ and
$W_{\beta_i}$, the consistency conditions for Wilson lines $W$ \ber :
\eqn\cons{MWM^{-1}=W^{-1}}
on $S^1/{\bf Z_2}$ and
\eqn\constwo{MWM^{-1}=W}
on $T^5$ imply that only $W_{\beta_i}$ can exist on $T^5$; whereas, only
$W_{\alpha_i}$  Wilson lines can exist on $S^1/{\bf Z_2}$.  This makes
sense since in the type I picture there is nothing on $T^5$ corresponding
to a $W_{\alpha_i}$  Wilson line, while $Sp(n)$ ($W_{\beta_i}$) Wilson
lines are not allowed on the transverse $S^1$.  Adding $W_{\alpha_i}$  
Wilson lines to the $S^1/{\bf Z_2}$ corresponds in the type I picture
to moving the fivebranes away from $\theta=0$.

In two dimensions we will have eight adjoint scalar multiplets of
$U(2n)$ consisting of an adjoint scalar and two adjoint fermions.
These eight multiplets come from the transverse spatial directions to
the $S^1/{\bf Z_2}$.  We can separate out the center of mass $U(1)$ and
count the number of center of mass states.  This center of mass
which commutes with the remaining $SU(2n)$ is
\eqn\cent{G_{\alpha}={1\over n}{\sum_{i=1}^n G_{\alpha_i}}.}
The monodromy acts as an overall ${\bf Z_2}$ on the scalar multiplets
coming from the $T^5$ but leaves the scalar multiplets corresponding
to $R^3$ invariant.  Thus, the number of center of mass states is
$(2^5/2)\times 2^3=16\times 8$, exactly what we obtained in the T-dual
picture of type I.  There is another $U(1)\in Sp(n)$ that has been
discussed previously, that is, 
$G_{\beta}={1\over n}{\sum_{i=1}^n G_{\beta_i}}$.  Giving scalars 
that correspond to $T^5$ Wilson lines $W_{\beta}^{\rho}$ generated
by $G_{\beta}$ expectation values breaks the $SU(2n)$ to
$U(1)\times SU(n)\times SU(n)$.  The monodromy exchanges and complex
conjugates the two $SU(n)$'s, and the $SU(n)$ subgroup of $Sp(n)$ is
generated by the difference of corresponding pairs of $SU(n)$ generators.

The next step is to see whether the left over $SU(2n)$ theory has 
a bound state.  As argued in Ref.~\nine\ , the
$N=8$ supersymmetry in two dimensions allows for a supersymmetric
vacuum with a mass gap.  The $m$ wound onebrane should be considered
to contribute a $SU(2n)$ electric flux away from the fixed points of the
the $S^1/{\bf Z_2}$ equivalent to
the antisymmetric tensor product of $m$ $SU(2n)$ fundamentals and 
a $U(1)$ flux of charge $m$.  We can now repeat the analysis of
Ref.~\nine .
The potential energy in the two dimensional gauge theory is
\eqn\pot{V=const\sum_{i,j=2}^9 Tr[X^i,X^j]^2}
where the $X^i$ are in the adjoint representation of $SU(n)$.
A supersymmetric quantum vacuum with $V=0$  and no charge on the boundaries
requires that $SU(n)$ be
broken to $U(1)^{n-1}$ by taking the eigenvalues of $X^i$ to be
large and distinct.  Classically, there are many other possibilities.
It was then shown that for any direction in
this vacuum, the electric charge would not be screened, and there would
be an energy barrier to making $X^i$ large if $n$ and $m$ were 
relatively prime.  The mass gap allowed the superpotential to be
perturbed without changing the ground state degeneracy.  The unique
solution that screened the charge spontaneously broke the $SU(n)$ gauge
symmetry causing all fields except the center of mass degrees of freedom 
to be massive.  

In our case the gauge symmetry is $SU(2n)$ with $n$ and $m$ relatively
prime, and we are on $S^1/\bf{Z_2}$.  
If $m$ is odd, then $2n$ and $m$ are relatively prime, and there is an
energy barrier in every direction to making the $X^i$ large.  This
energy barrier allows one to to perturb the superpotential of
the $N=4$ supersymmetric Yang-Mills theory in four dimensions (this term
reduces unchanged to two dimensions where there are additional terms)
written in terms of $N=1$ fields,
\eqn\sup{S=Tr(A[B,C]),}
without changing the ground state degeneracy where $A$, $B$, and $C$
are the three chiral superfields in the adjoint representation of
$U(2n)$.  A vacuum state requires that these fields form a $2n$
dimensional representation of $SU(2)$.  There is a unique supersymmetric
vacuum state in which the $SU(2n)$ gauge symmetry is spontaneously
broken and the boundary charge screened corresponding to the
case that the representation is irreducible.  By taking the radius of
$S^1/\bf{Z_2}$ to infinity, we recover the situation of Ref.~\nine\ ,
and the bound state does not disappear.
If $m$ is even, we must ask whether screening occurs when we break
$SU(2n)$ to $U(1)\times (SU(n))^2$ by varying the $T^5$ $G_{\beta}$
scalars.  Here the charges of one of the fundamental $n$'s of $SU(n)$
with respect to $G_{\beta}$ will all be $+1$, and the charges of the 
other $n$ will all be $-1$.  Since we know that there are $(n,{m\over 2})$
bound states by the above argument, there would appear to be no
energy barrier in this direction.  However, the monodromy exchanges
the two bound states leaving a unique bound state with an energy
barrier to separation in this direction.  We conclude that there is
a unique bound state whether $m$ is odd or even.  As noted in 
Ref.~\ten\ the conclusion should not change when some of the
$X^i$ need to be periodically identified (as they do here because
we are dimensionally reducing on $T^5$ not $R^5$) because the 
bound state is localized in field space.  For the same reason, the 
conclusion should also be valid away from the singularities where
the supersymmetry on the brane is halved, and charged hypermultiplets
become massless.  Since turning on the
$Sp(n)$ Wilson lines in the unreduced theory is equivalent to
giving expectation values to the periodic scalars in the dimensionally
reduced theory, there must be an energy barrier to activating
these lines in the unreduced theory.
We have, therefore, found $16\times 8$ of the H-dyon
states as in Ref.~\six\ but one-half the number of Ref.~\ten\ because
of the ${\bf Z_2}$ action.

\subsec{Singular H-Dyons}

Let us try to find the other expected $16\times 16$ H-dyon states
with $n,m$ relatively prime.  We have seen that in the type I' picture
there is generically an energy barrier to turning on the $W_{\beta_i}^{\rho}$
Wilson lines on $T^5$.  However, there are special points in the space of
$W_{\beta_i}^{\rho}$ where the interaction with $SO(32)$ Wilson lines
cause charged fundamentals of the gauge group to become massless.  At
these points the generic $N=2$ supersymmetry in six dimensions will be
halved.  If we T-dualize on $T^5$ at these points, we find that at least
one onebrane will touch a threebrane there.  To obtain the expected
H-dyon states, we expect that in general cases there will be no supersymmetric
ground states unless all the $W_{\beta_i}^{\rho}$ are adjusted so that an
extra $2n$ charged hypermultiplets (in the six dimensional sense of
hypermultiplet) become massless.  In the case that $n$ and $m$ are
relatively prime, we require that the $W_{\beta_i}^{\rho}$ are such that
$\beta_i=\beta^0$, all $i$ where $\beta^0$ is determined by a $U(1)$
Wilson line of $SO(32)$.  Consider $m$ odd and $(n,m)$ relatively
prime.  Fixing the $W_{\beta_i}^{\rho}$, we know that every direction 
varying the $W_{\alpha_i}$ lines in $S^1/{\bf Z_2}$ or the $R^3$ scalars
contains an unbroken $U(1)$ that is charged with respect to every
component of the charge on the boundary.  Unless all of the nonabelian flux
is screened, there is an energy barrier in some of these directions.  
Suppose we could separate the system into subsystems by moving along
directions where the nonabelian flux is screened.  The resulting vacuum
states could not be supersymmetric because there would be an 
energy barrier in all the other
directions which is not permitted by the $N=4$ supersymmetry in two
dimensions \nine .  If $m$ is even the above possible vacuum state
is split into two; but because of the ${\bf Z_2}$ monodromy, we
also find an energy barrier in all other directions for this case.
Thus, we require $2n$ charged states to become massless so that all of the
nonabelian flux can be screened.

We now address the question of why only singularities with all the
$W_{\beta_i}^{\rho}$ having $\beta_i=\beta^0$ should give supersymmetric
ground states in the relatively prime case.  From the above discussion
it is clear that there will be an energy cost to varying any of the 
$W_{\beta_i}^{\rho}$ away from the singularity because some of the $2n$
hypermultiplets will gain a mass.  However, we might imagine that there
will be another minimum of the potential with several subsystems
$(2n^s, m^s)$ with $n^s$, $m^s$ relatively prime such that $2n=2\sum_s n^s$
charged hypermultiplets become massless at this singularity.  This
cannot occur in the relatively prime case for the following reason.  
Not all of these subsystems will be identical so that the permutation
symmetry will be violated, and we require that a supersymmetric ground
state be invariant under the Weyl symmetry.  In the case that $n$ and
$m$ are not relatively prime, there can be such subsystems which respect the 
permutation symmetry.

There are sixteen singularities that meet the above requirements for $n,m$
relatively prime.  They correspond to adjusting the $W_{\beta_i}^{\rho}$ 
Wilson lines to interact with each of the sixteen $U(1)$'s of $SO(32)$
such that in T-dualizing on $T^5$ we find $n$ onebranes touching one of
the sixteen threebranes at each singularity.  Now we need to argue
that each of these singularities gives sixteen bound states.  
At the singularity the gauge symmetry is $U(1)\times U(1)\times (SU(n))^2$
where the two $U(1)$'s are $G_{\alpha}$ and $G_{\beta}$, and the 
monodromy exchanges the two $SU(n)$'s.  In addition to the adjoint
scalars we have an $(n,1)+(1,n)$ of $(SU(n))^2$ corresponding to an
$N=2$ hypermultiplet in four dimensions.  Since all of the boundary
charge can be screened by the charged fundamentals, a vacuum 
solution requires breaking the gauge symmetry to $U(1)^{2n}$ by
giving expectation values to the adjoint scalars parametrizing $R^3$
and the $W_{\alpha_i}$ Wilson lines in the $S^1$.  These fields do
not interact with the charged hypermultiplets.  Remember that we 
can only give expectation values to the neutral scalars in the diagonal
$SU(n)$ of $SU(n)^2$ because of the monodromy.  Including the center
of mass scalars, we see that the nonabelian gauge symmetry can only
be broken by assigning {\it distinct} expectation values to the 
$n$ neutral ``hypermultiplets''.  By hypermultiplets I mean the four
scalars parametrizing $R^3\times S^1$.

We now have to argue that each such configuration of hypermultiplets
gives a unique supersymmetric ground state modulo a $U(n)$ gauge
transformation.  (Any diagonal $U(n)$ transformation will preserve
the eigenvalues of the adjoint scalars.)  Our argument will be 
similar to that of Ref.~\eleven .  In four dimensions, 
the $N=2$ supersymmetric
theory has the following coupling between the the neutral 
chiral multiplets of the $U(1)$ vector multiplets parametrizing
two Wilson lines on $T^5$ and the charged chiral multiplets composing the
charged hypermultiplets:
\eqn\coup{S=\sum_{i=1}^n ({\bar\Lambda}_i^1 (\phi_{\beta_i} -
\phi_{\beta^0})\Lambda_i^1
-{\bar\Lambda}_i^2 (\phi_{\beta_i} -\phi_{\beta^0})\Lambda_i^2)}
where $\phi_{\beta^0}$ is the value at the singularity.  The 
$\phi_{\alpha_i}$'s do not contribute because of the monodromy.
We seem to be singling out two directions in $T^5/{\bf Z_2}$ (the
space of $W_{\beta_i}^{\rho}$ with $i$ fixed), but the $SL(5)$ symmetry
of $T^5/{\bf Z_2}$ allows us to always rotate the coupling to this form.
This coupling should reduce unchanged to two dimensions.  Again,
the neutral hypermultiplets are uncoupled to these charged fields.
There is an energy barrier to perturbing $\phi_{\beta_i}$ because 
the charged fields
gain a mass and there is an unbroken $U(1)$ flux that is unscreened as 
in our previous discussion.  Thus,
$S$ can be perturbed by adding some other terms \eleven\ , and taking into
account the monodromy one
finds a unique solution, modulo a $U(1)^n$ transformation, that breaks
the $U(1)$'s causing all fields except the neutral hypermultiplets 
to be massive and screens the electric flux.
Thus, the space of solutions is determined by the $n$ neutral
hypermultiplets modulo a gauge transformation.  By finding this solution
that screens the flux, we have effectively reduced the system to the
$N=4$ supersymmetric quantum mechanics on the moduli space of hypermultiplets.

Let us return to the heterotic string picture.
The hypermultiplets have an interpretation as the 
location of small instantons in $R^3\times S^1$.  These instantons
have electric charge determined by the momentum on $S^1$.  The moduli
space of hypermultiplets is, thus, equivalent to the 
desingularization of $S^n(R^3\times S^1)$($S^n$ is the symmetric
product of $n$ elements).  It is argued in Ref.~\twelve\
that this moduli space is diffeomorphic to that 
of the BPS $n$-monopole space.
Since noncompact spaces generally have a choice of 
desingularizations, and we would not want to treat this space as
an orbifold in string theory (an approach which would give the wrong
answer), we assume that the Hilbert scheme approach of Ref.~\twelve\ is 
the correct one.
Allowing for the action of the charge generator on this space,
we obtain the moduli space of BPS $(n,m)$-dyons where we have assumed
that whether the charge is from winding or momentum does not
affect the moduli space.  The center of mass can be factored out
giving the usual sixteen states.  At this point it is natural to
conjecture that $n$ small instantons with momentum $m$ ($n$, $m$ 
relatively prime) at such
a singularity where the nonperturbative gauge symmetry is eaten
are the same as a BPS $(n,m)$-dyon.
Thus, assuming the conjecture
of Ref.~\five\ is correct, we get the extra $16\times 16$ states
from the
 singularities, raising the total to $16\times 24$ as expected.
Note that it is claimed in Ref.~\por\ that
the conjecture of Ref.~\five\ has been proven, and evidence for
the conjecture has also been presented in Ref.~\seg .

Let us try to make our conjecture more plausible.  As in our previous 
discussion, at the singularity we will have a common $U(1)$ shared by
$SO(32)$ and the overall $Sp(1)$ of the center of mass.  
There will be no nontrivial dependence on $T^6$
except for the momentum on the $S^1$ (Assume that we are in the type I
picture.)  All nonperturbative gauge symmetry 
is broken so we expect a nonsingular
solution.  The mass of the resulting state is related to the expectation
values of the overall Wilson lines $W^{\rho}_{\beta}$.  Half 
the supersymmetry on the
D5 brane is broken at the singularity with the appearance of a BPS dyon.
Finally, the moduli space of the small instantons at the singularity
agrees with that of
the BPS dyon.

\newsec{Conclusions}

We have located the $16\times 24$ $(n,m)$ H-dyon states with $n,m$
relatively prime expected by S-duality of the heterotic string. 
Our results are valid at all points of moduli space where the 
perturbative gauge symmetry is abelian.  Enhanced perturbative gauge
symmetry should not change this picture for the $16\times 8$ states
at generic points of the moduli space, but the $16\times 16$ states
can only come from small instantons.
We have conjectured and found evidence that at 
singularities where the nonperturbative
gauge symmetry is completely broken, the H-dyon bound
state should be equivalent to a BPS dyon bound state.  

The elementary heterotic states on $T^4\times S^1_5\times S^1_6$ with
$N_L=1$ that have momentum on $S^1_6$ correspond to Type IIA states
on $K3\times S^1_5\times S^1_6$
with momentum on $S^1_6$.  The heterotic states that are magnetically
charged with respect to $B_{\mu 6}$ ($B$ is the antisymmetric tensor)
turn into winding states on $S^1_5$ that are electrically charged
with respect to $B_{\mu 5}$.  The mass of the $(n,m)$ states in
Type IIA is 
\eqn\mtwo{M^2\sim{m^2\over R^2_6}+{n^2\times R_5^2\over\alpha^{'2}}}
so that if $(n,m)$ are relatively prime, these states are stable.
Either by reducing the low energy IIA supergravity on $K3\times T^2$
or by calculating directly the IIA theory on a K3 orbifold, one
obtains $24\times 16$ states with oscillators in their ground states.
Thus, the total number of H-dyon states in the 
Type IIA picture with $(n,m)$ relatively prime is $24\times 16$, and
our results provide further evidence in favor of the string-string
duality conjectures in six dimensions.  All of our arguments have required
$n$ and $m$ to be relatively prime.  In other cases, there are 
marginally bound states \dh\ that produce degeneracies that are not
simply the BPS result multiplied by $24$, and we do not expect our
conjecture to be valid in these cases.

\bigskip

I wish to thank K. Dienes, K. Intriligator, S. Sethi, and especially E. Witten
for useful discussions.
This research was supported in part by NSF grant PHY-9513835.

\listrefs
\end